\documentclass[preprint,3p,12pt]{elsarticle}
\biboptions{sort&compress,comma}

\usepackage{braket}

\usepackage[unicode, pdfencoding=auto]{hyperref}

\usepackage[utf8]{inputenc}
\usepackage{csquotes}
\usepackage{amssymb}
\usepackage{amsmath}

\usepackage[numbers]{natbib}

\journal{Journal of Alloys and Compounds}

\begin{document}

\begin{frontmatter}

\title{Tailoring hard magnetic properties of Fe$_2$MnSn Heusler alloy via interstitial modification: A first-principles approach} 

\author[1]{Junaid Jami} 
\ead{junaid.jami777@gmail.com}

 \author[2]{Rohit Pathak} 
\ead{rohit.pathak@physics.uu.se}

\author[3]{N. Venkataramani} 
\ead{ramani@iitb.ac.in}

\author[4]{K.G. Suresh} 
\ead{suresh@iitb.ac.in}

\author[1]{Amrita Bhattacharya\corref{cor1}}

\ead{b_amrita@iitb.ac.in}

\cortext[cor1]{Corresponding author}

\affiliation[1]{organization={AbCMS Lab, Department of Metallurgical Engineering and Materials Science, Indian Institute of Technology Bombay},
            addressline={Powai}, 
            city={Mumbai},
            postcode={400076}, 
            state={Maharashtra},
            country={India}}
\affiliation[2]{organization={Materials Theory Division, Department of Physics and Astronomy, Uppsala University},
            city={Uppsala},
            postcode={751 20}, 
            country={Sweden}}
\affiliation[3]{organization={Department of Metallurgical Engineering and Materials Science, Indian Institute of Technology Bombay},
            addressline={Powai}, 
             city={Mumbai},
            postcode={400076}, 
            state={Maharashtra},
            country={India}}
\affiliation[4]{organization={Department of Physics, Indian Institute of Technology Bombay},
            addressline={Powai}, 
             city={Mumbai},
            postcode={400076}, 
            state={Maharashtra},
            country={India}}

\begin{abstract}

We employ first-principles calculations to explore interstitial engineering as a strategy to tailor the hard magnetic properties of Fe$_2$MnSn Heusler alloy, establishing its potential as a rare-earth-free permanent magnet. By introducing light interstitial elements—B, C, H, N, O, and F—at varying concentrations (1.56–12.5 at\%), we uncover significant enhancements in structural stability, magnetization, Curie temperature, and magnetocrystalline anisotropy. These dopants preferentially occupy octahedral interstitial sites in the hexagonal phase of Fe$_2$MnSn, leading to localized lattice distortions that enhance its magnetic characteristics. Notably, at 12.5 at\% doping, B, C, N, and O induce a critical transition from in-plane to out-of-plane magnetic anisotropy—achieved without 5$d$ or rare-earth elements—highlighting a sustainable pathway to high-performance magnets. Among these, N-doped Fe$_2$MnSn exhibits the highest uniaxial anisotropy (0.61 MJ/m$^3$), followed by B-doped (0.44 MJ/m$^3$) alloy. The magnetization of the doped compounds surpasses that of conventional ferrites and gap magnets like MnAl and MnBi. The Curie temperature sees a substantial boost, reaching 1058 K for O-doped Fe$_2$MnSn and 1000 K for C-doped alloy. Although N-doping results in a modest increase in T$_{\mathrm{C}}$ (744 K vs. 729 K for the pristine alloy), it delivers superior hard magnetic properties, with the highest magnetic hardness (0.65) and an enhanced maximum energy product (0.36 MJ/m$^3$), making it a strong candidate for gap magnet applications. These findings highlight interstitial doping as a viable route to engineer rare-earth-free permanent magnets with optimized magnetic performance.

\end{abstract}

\end{frontmatter}



\section{Introduction}
\label{sec1}
The growing concerns about climate change and the push to achieve net-zero greenhouse-gas emissions have accelerated the global search for alternatives to traditional fossil fuel–based energy sources and  efficient energy utilization \cite{RIZOS20241673,BARRETO2018196}. The global transition toward renewable energy and a low-carbon economy is a compelling trend, driven by emerging technologies that rely heavily on permanent magnets (PMs) \cite{MOHAPATRA20181,TRENCH2020115,CUI2018118}. Large volumes of PMs are used in hybrid and electric vehicles \cite{WIDMER20157}, direct-drive wind turbines \cite{PAVEL2017349}, and energy-efficient appliances \cite{mccallum2014practical}.    Current PM market is split between low performing yet cost-effective ferrites and high performing but expensive rare-earth (RE) magnets primarily based on Nd-Fe-B and Sm-Co \cite{doi:10.1021/acssuschemeng.3c02984,8002189}. However, concerns over resource exhaustion and the limited global supply of RE elements — coupled with significant economic and environmental costs, have prompted extensive research efforts aimed at developing high-performance RE-free PMs or magnets that will bridge the performance gap between RE-based permanent magnets and ferrites, commonly referred to as gap magnets \cite{stegen2015heavy,coey2020perspective,skomski2013predicting,lee2020anisotropic}. Materials design of new magnets requires tailoring and controlling the intrinsic magnetic properties of the material, with special emphasis on magnetization, Curie temperature (T$_{\mathrm{C}}$) and uniaxial magnetic anisotropy.

Computational materials design using \emph{ab initio} methods or atomistic modeling offers a resource-efficient and often quicker alternative to experimental approaches. Broadly, two primary strategies exist: high-throughput data mining to discover new, unknown phases,\cite{MAL2024171590, vishina2020high, vishina2021data} or the optimization and modification of known structures,\cite{kan2016tuning, PhysRevB.98.014411}  where the main challenge lies in identifying the key variables that must be adjusted to enhance the relevant properties. Lately, there has been growing interest in transition metal based permanent magnetic materials, particularly Heusler alloys, as they may exhibit a range of viable functional applications, and ease in tuning their electronic and magnetic properties \cite{SKOKOV2018289,GAO2020355,TAVARES2023101017}. Moreover, their extensive combinatorial possibilities significantly increase the likelihood of discovering new magnetic materials \cite{wederni2024crystal, graf2011simple}. Special focus is on Fe-based Heusler alloys with low-symmetry (non-cubic) structures as promising candidates for permanent magnets. This focus is driven by the  abundant iron resources in the Earth's crust, the sizable magnetic moment and relatively high Curie temperature offered by Fe atoms, and most importantly, the presence of uniaxial magnetic anisotropy \cite{LI2019535, faleev2017heusler, MENG2019224, PhysRevB.87.064411}. The latter being an essential factor for achieving higher coercivity and enhancing the maximum energy product BH$_{\mathrm{max}}$. 

Since the theoretical prediction of Fe$_2$MnSn (FMS) exhibiting a large magnetic moment and high T$_{\mathrm{C}}$, extensive theoretical studies, experimental synthesis, and characterization have been undertaken to explore its compelling magnetic properties \cite{faleev2017heusler, faleev2017origin, dahal2020electronic, KRATOCHVILOVA2020166426}.  In our previous work \cite{PhysRevB.108.054431}, we have shown that the stable hexagonal D0$_{19}$ ($P6_3/mmc$, space group no. 194) phase of FMS  exhibits a ferromagnetic ordering with large magnetization  6.45 $\mu_\mathrm{B}$/f.u. and a high T$_{\mathrm{C}}$ 729 K. For PM applications, it is desirable to stabilize the local magnetization direction to provide a uniaxial magnetic state which is often achieved by developing magnetic anisotropy \cite{Hirosawa_2017}, which was lacking in FMS. In this work, we adopt the latter approach of computational materials design focused on the modification and optimization of known structure (FMS) and have tried to engineer an out-of-plane magnetic anisotropy, thereby enhancing its potential for application as a permanent magnet.

The shape and orientation of $d$-orbitals, which are highly sensitive to structural changes and the chemical environment of the metal atoms, significantly influence the magnetic anisotropy energy (MAE), making it responsive to modifications in the material's composition and electronic structure \cite{PhysRevX.4.021027,PhysRevB.98.014411}.  One potential route to tailor the magnetic properties of transition metal alloys is interstitial modification, which can alter crystallinity, local atomic arrangements, symmetry (space group) and the distribution of structural and electronic defects \cite{xu2024interstitials, KHAN201717, PhysRevApplied.11.054085, GAO2020355}, leading to alterations in the density of states near the Fermi level \cite{kitagawa2020interstitial,islam2021large}. The formation of interstitial alloys requires that the foreign species (typically from the first period of the $p$-block and below) be small enough to diffuse into the interstitial sites. These light elements are characterized by small atomic radii and dominant $s–p$ orbitals, and often occupy the largest available interstitial site \cite{chen2021interstitial}.

\begin{figure}[t]
\centering
\includegraphics[scale=0.26]{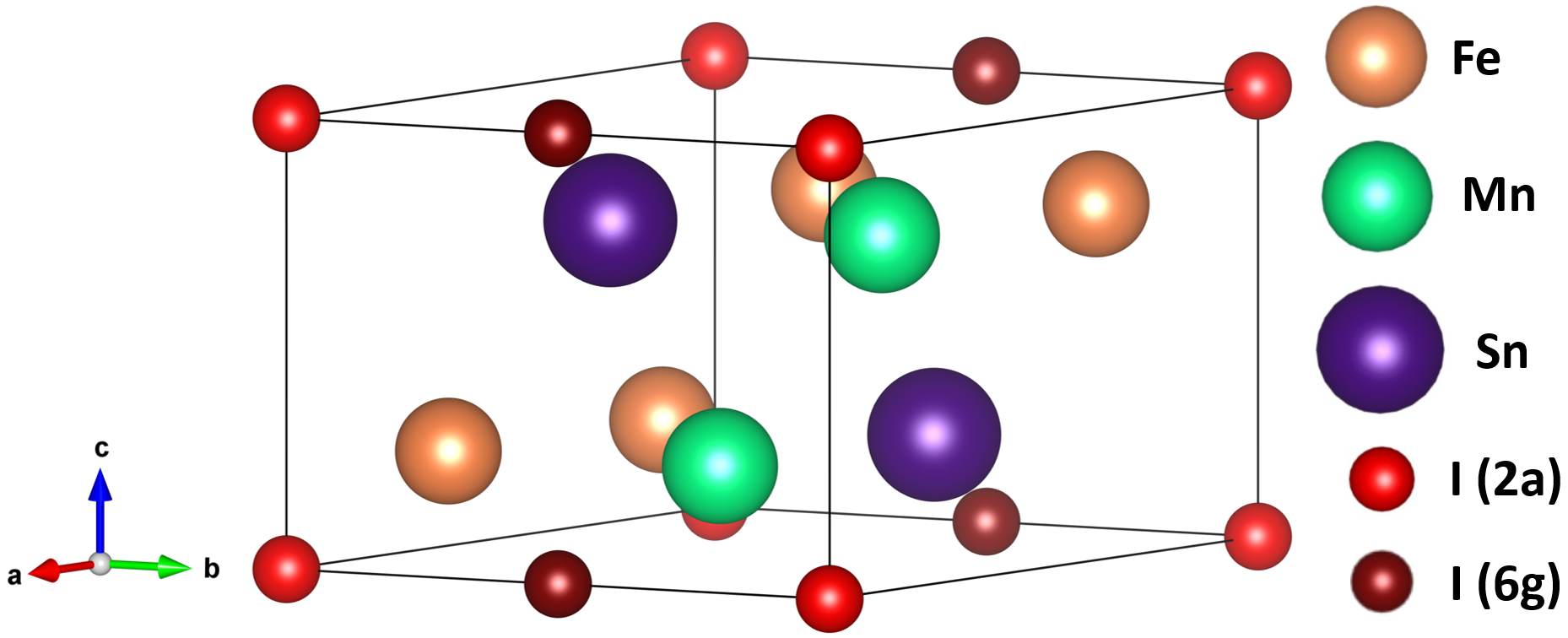}
\caption{The conventional unit cell of interstitially modified Fe$_2$MnSn Heusler alloy with the interstitial atom (I) occupying the octahedral $2a$ (0,0,0) and $6g$ (0,0.5,0) Wyckoff sites.}
\label{figure:1}
\end{figure}

Gao \emph {et al.} \cite{GAO2020355} conducted high-throughput first-principles calculations to investigate the effects of doping 6.25 at\% light interstitial atoms (B, C, H, and N) in 128 cubic Heusler alloys, aiming to design new RE-free PMs. They found that octahedral centers were the  preferred interstitial sites for most compounds, and doping induced tetragonal distortion in the cubic systems, leading to an uniaxial MAE in as many as 32 compounds (MAE $>$ 0.4 MJ/m$^{3}$). The addition of these small atoms expanded the lattice parameters, increasing Fe–Fe interatomic distances (for Fe-based Heusler alloys) thereby enhancing the T$_{\mathrm{C}}$.  In their high-throughput calculations, Opahle \emph {et al.} \cite{PhysRevResearch.2.023134} demonstrated that the interstitial B, C, and N atoms in 21 cubic Cu$_3$Au-type structures significantly influence not only the MAE but also the thermodynamic stability, the magnetic ground-state structure $-$ aligning with the findings of several other studies \cite{zhang2016engineering,gajdzik2000ferromagnetism,trung2010giant}.  Interstitial modification has also been reported to greatly enhance the strength and rigidity of (Ni$_{2}$Mn$_{1.5}$Ti$_{0.5}$)$_{1-x}$B$_{x}$ ($x$ = 0.03, 0.06, 0.09), as a small part of the $d$-$d$ hybridization in ternary Ni-Mn-Ti alloy is replaced by the $p$-$d$ hybridization in Ni-Mn-Ti-B alloy \cite{xiong2022first}. B-doped Ni$_{36.5}$Co$_{13.5}$Mn$_{35}$Ti$_{15}$ alloys exhibited a 40\% increase in saturation magnetization \cite{PhysRevApplied.17.054032}, while interstitial oxygen and nitrogen enhanced the MAE and T$_{\mathrm{C}}$ in CoNiPt$_2$ and CoFePt$_2$ alloys, respectively \cite{ALEDEALAT2023111649}. Luo \emph {et al.} \cite{LUO2020106971} synthesized  Fe$_2$MnGaC$_x$ ($x$ = 0, 0.05, 0.1 and 0.2) alloys and reported carbon doping  makes the $d$-states of the magnetic atoms (Fe and Mn) more localized thereby increasing the partial spin moments. Moreover, the introduction of C suppressed the ferromagnetic to antiferromagnetic transition in Fe$_2$MnGa,  enhancing both magnetization and T$_{\mathrm{C}}$. Similarly, Zhang \emph {et al.} \cite{ZHANG201426} demonstrated that by precisely adjusting the C content in cubic L2$_1$ Ni-Mn-Sn Heusler alloy the magneto-caloric properties could be tailored, yielding high magnetic entropy changes and increased compressive strength. Meng \emph{et al.} \cite{MENG2020167161} showed that B doping in cubic L2$_1$ Fe$_2$MnSi Heusler alloy enhances ferromagnetism, with T$_{\mathrm{C}}$ rising from below room temperature to about 400 K for $x = 0.3$. Overall, the interstitial light elements play an important role in  materials by improving the magnetic properties through changes of the unit cell volume or orbital hybridization between the magnetic and interstitial atoms.
 
 
 In the present study, using calculations based on density functional theory, we propose doping with light interstitial atoms as an effective tool to engineer out-of-plane magnetic anisotropy and magnetic properties in Fe$_2$MnSn Heusler alloy.
 Consequently, inspired by the effect of doping with light interstitials in Heusler alloys, in the present study we have tried to successfully engineer an out-of-plane magnetic anisotropy in FMS and explore the enhacement of its magnetic properties through interstitial modification using DFT, with the aim to explore the potential applications of FMS as a gap-magnet.
 We selected six different interstitial atoms (B, C, H, N, O, and F) and have varied the doping concentrations to 1.56, 3.125, 6.25 and 12.5 at\%. Significantly, at 12.5 at\%, all the interstitially doped compounds except H and F-doped FMS show a switch from in-plane to out-of-plane anisotropy, with N doped-FMS (N-FMS) showing the highest MAE. Additionally, most of the doped compounds display enhanced T$_{\mathrm{C}}$ compared to the undoped FMS. An increase in the net magnetic moment of FMS is also realized for all dopants except in the cases of B and C. This work demonstrates that interstitial atoms play a crucial role in stabilizing the crystal structure and modifying the magnetic properties of FMS thus leading to the inference that interstitial doping can be a vital tool for the applications of these alloys as RE-free permanent magnet.

\section{Computational Details}
\label{sec2}
An in-depth analysis, to identify at first energetically favorable interstitial sites and then to study the effects of interstitial dopants (B, C, H, N, O, and F) on the electronic and magnetic properties of Fe$_2$MnSn  alloy was performed in the framework of DFT \cite{PhysRev.136.B864, PhysRev.140.A1133} using Vienna $ab$ $initio$ simulation package (VASP)\cite{PhysRevB.54.11169}. The exchange and correlation effects between the electrons were treated by employing the projector-augmented wave (PAW) potentials \cite{PhysRevB.50.17953, PhysRevB.59.1758} using the parametrization given by Perdew, Burke, and Ernzerhof within the generalized gradient approximation (GGA) \cite{PhysRevLett.77.3865}. 3$d$ and 4$s$ electrons serve as valence electrons for Fe and Mn, while for Sn 5$s$ and 5$p$ are set as valence electrons. Spin-polarized calculations were carried out for atomic as well as geometric relaxation. To ensure the convergence and reliability of our results, the plane-wave cut-off energy was set to 550 eV, while the k-point mesh was chosen according to the Monkhorst-Pack \cite{monkhorst1976special} scheme to achieve sufficient Brillouin zone sampling. The lattice structures were optimized with the energy uncertainty of 10$^{-5}$ eV  and relaxed by restricting the interatomic forces below 0.001 eV/\AA. For visualization of crystal structure,  VESTA \cite{momma2011vesta} was utilized.


 For the interstitially modified compounds, the formation energy ($E_\mathrm{f}$) at 0 K is calculated by taking the difference between the total energy of the doped compound \emph{E}$_\mathrm{doped}$ and the undoped (pure) \emph{E}$_\mathrm{undoped}$ compound added with the chemical potential of the interstitial atom $\mu_{I}$.
 \begin{equation}
 	\label{eqn:E1}
 {E_\mathrm{f} = E_\mathrm{doped} - E_\mathrm{undoped} + \mu_{I}} 
\end{equation}

The diatomic molecular form (H$_2$, O$_2$, N$_2$ and F$_2$) is considered as the reference state for H, O, N and F respectively, while the diamond carbon and trigonal structure are considered as reference states for C and B, respectively.

MAE is a crucial intrinsic property that contributes to the performance of permanent magnetic materials and is defined as the energy penalty incurred from changing the  magnetization orientation from one direction (in-plane) to another direction (out-of-plane) within a crystal.  As proposed by Van Vleck \cite{PhysRev.52.1178}, the contribution of itinerant ferromagnetic electrons to MAE arises from the spin-orbit interaction that couples the spin and orbital components of the magnetic moments. We have calculated MAE using the magnetic force theorem \cite{wang1996validity} by taking the difference of the total energies for magnetization directions [100] and [001], 
\begin{equation}
	\label{eqn:E2}
	\mathrm{MAE}=E^{[100]}-E^{[001]}
\end{equation}
where $E^{[100]}$ and $E^{[001]}$ are the total energies of the material with magnetization aligned along the hard and easy axis, respectively. The global spin quantization axis is varied along the three crystallographic orientations i.e., the a, b and c directions, and the total energies are compared for each case. MAE is calculated as the difference between the total energy of the magnetization oriented along the in-plane ($E^{[100]}$) and out-of-plane directions ($E^{[001]}$).
The magnetic anisotropy energy density or the first order anisotropy constant ($K$) can then be calculated as the MAE per unit volume $(V)$. 
\begin{equation}
	\label{eqn:E3}
	K=\frac{\mathrm{MAE}}{V}
\end{equation}
 A positive value of MAE or $K$ corresponds to an uniaxial anisotropy which means the easy magnetization axis is perpendicular to the $ab$-plane i.e., along the $c$-axis, indicating the material's potential for permanent magnet applications. After identifying the appropriate dopants and doping concentration (12.5 at\%) required to engineer uniaxial anisotropy in FMS, further calculations were performed for these compounds at the specified interstitial concentration. We analyze the percentage of spin polarization (SP) \cite{soulen1998measuring} using the popular definition:

\begin{equation}
	\label{eqn:E4}
\mathrm{SP}=\left(\frac{\mathrm{\eta}^{\uparrow}-\mathrm{\eta}^{\downarrow}}{\mathrm{\eta}^{\uparrow}+\mathrm{\eta}^{\downarrow}}\right)_{\mathrm{E_F}} \times 100 ~ \%
\end{equation}       

where, SP is the percentage of the ratio of difference in the density of majority spin (${\eta}^{\uparrow}$) and minority spin (${\eta}^{\downarrow}$) states and the total density of spin states at the Fermi level E$_\mathrm{F}$. 


The Heisenberg exchange coupling parameters $J_{ij}$ were calculated by mapping the system onto a Heisenberg Hamiltonian, using the method proposed by Liechtenstein \emph {et al.} \cite{liechtenstein1987local}, as implemented in spin-polarized relativistic Korringa-Kohn-Rostoker package, Munich SPRKKR \cite{ebert2011calculating}. The calculations were carried out within the scalar relativistic mode and for consistency, the same functional for exchange and correlation has been chosen as before \cite{PhysRevLett.77.3865}.  The angular momentum expansion up to $l_{max}$ = 3 has been taken for each atom to ensure the accurate description of the electronic structure near the atomic sites and the integration over the Green’s functions was carried out with 30 energy points. In order to further improve the charge convergence with respect to $l_{max}$,  Lloyd’s formula has been employed for determining the Fermi energy \cite{doi:10.1080/00018737200101268, Zeller_2008}. The convergence criterion was set at level of $10^{-5}$ and 10,000 k-points were employed per reduced Brillouin zone. The exchange coupling parameters are calculated with respect to the central site $i$ for cluster atoms with radius $R_{clu} = \mathrm{max}|R_i-R_j|$. We have taken the radius of a sphere $R_{clu}$ of 7.0. The Curie temperature, is then calculated within the mean-field approximation using the approach discussed in \cite{PhysRevB.108.054431}.




Bonding analysis was carried out by studying crystal orbital Hamilton population (COHP)\cite{dronskowski1993crystal} based on band energy partitioning. The wavefunctions of the simulated structures were generated by VASP and COHP analyses carried out via the LOBSTER program \cite{nelson2020lobster}, which projects the delocalized, plane-wave based information onto local orbitals. The integral of the COHP population curve up to the Fermi level (ICOHP) sums all bonding and antibonding interactions and hints towards bond strength \cite{steinberg2018crystal}.  

\section{Results}
\label{sec3}

\subsection{Site preference and stability}
\label{subsec1}

\begin{figure}[t]
\centering
\includegraphics[scale=0.50]{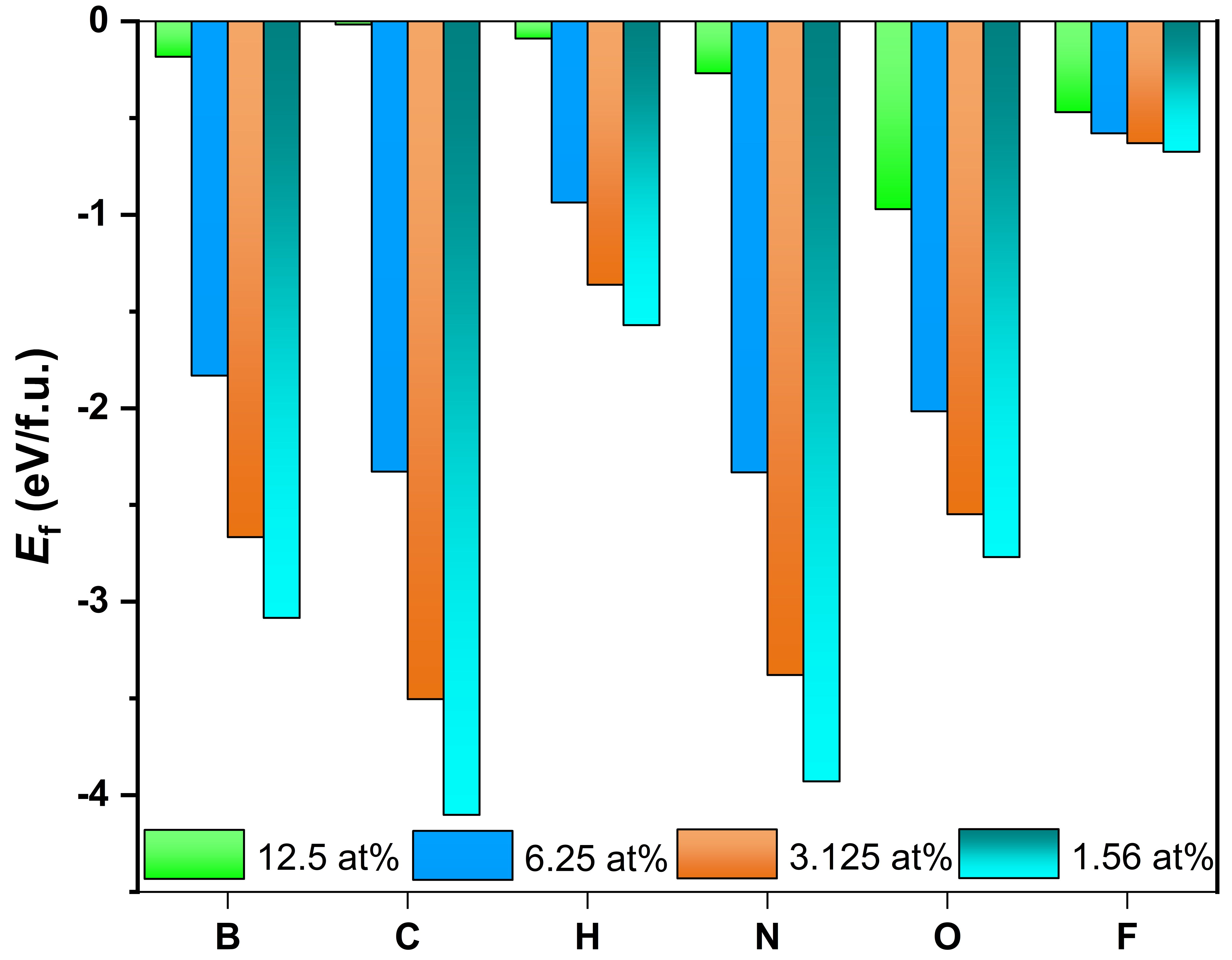}
\caption{Formation energy ($E_\mathrm{f}$) of interstitially modified Fe$_2$MnSn for varying interstitial concentrations viz. 1.56 at\%, 3.125 at\%, 6.25 at\% and 12.5 at\% with different interstitial atoms B, C, H, N, O and F. The $E_\mathrm{f}$ is found to be negative in all cases, while it is found to be less negative for higher interstitial concentration.}
\label{figure:2}
\end{figure}

In the hexagonal crystal structure of Fe$_2$MnSn \cite{PhysRevB.108.054431,KRATOCHVILOVA2020166426}, as shown in Fig. \ref{figure:1}, the Fe and Mn atoms occupy the $6h$ ($\frac{1}{6}$,$\frac{1}{3}$,$\frac{1}{4}$) sites with $2/3$ and $1/3$ occupancy, respectively, and the Sn atoms reside at the $2c$ ($\frac{1}{3}$,$\frac{1}{3}$,$\frac{1}{4}$) sites. To begin, an in-depth analysis of the  interstitial sites within the hexagonal Fe$_2$MnSn unit cell was conducted to determine the preferred positions of the interstitial atoms, namely boron (B), carbon (C), hydrogen (H), nitrogen (N), oxygen (O) and fluorine (F), chosen due to their smaller atomic radii. We can now refer to Fe$_2$MnSn doped with interstitial atoms as I-FMS, with \enquote{I} denoting the specific interstitial atom. For all the possible structures, geometric relaxation is performed while accounting the spin polarization to incorporate the effect stemming from the magnetic interactions, since undoped Fe$_2$MnSn is a known ferromagnet.

A thorough examination of the D0$_{19}$ system revealed multiple potential interstitial sites viz. the $2a$, $2b$, $2d$, $4e$, $4f$, $6g$, $6h$, $12i$, $12j$ and $12k$ sites \cite{aroyo2006bilbao,CONNETABLE201932307}.  The tetrahedral sites correspond to the $12k$ and $4f$ sites, while the octahedral sites correspond to the $2a$ and $6g$ sites, as corroborated by other studies \cite{bakulin2017absorption,wei2010site}.  The remaining sites are located on the faces or the edges of the octahedron or tetrahedron volumes and have been previously identified as potential insertion sites for FCC and HCP phases \cite{PhysRevLett.107.045504, scotti2016interstitial, david2017diffusion}, making them relevant for consideration in this study.  The interstitial concentration was varied using 2$\times$2$\times$2, 2$\times$2$\times$1, 2$\times$1$\times$1, and 1$\times$1$\times$1  supercells with one interstitial position occupied, corresponding to interstitial concentrations of 1.56, 3.125, 6.25, and 12.5 at\%. We limit the doping concentration to 12.5 at\% because higher interstitial concentrations can lead to excessive lattice distortion and potential phase instability, which could negatively impact the material's structural integrity. Additionally, concentrations around 12.5 at\% have been explored in both experimental and theoretical studies, \cite{sakurai2020metastable,zhang2016engineering,reichel2015soft,gutfleisch2011magnetic} offering a practical balance between enhanced magnetic properties and maintaining the stability of the crystal structure and making the chosen range as a realistic choice for our investigation. The preferred interstitial site was determined by comparing the formation energies of structures with interstitial atoms positioned at various available crystallographic sites. Our calculations reveal that the octahedral $2a(0,0,0)$ site is the most energetically favorable one for all dopants, except for fluorine (F), which prefers the octahedral $6g(0,0.5,0)$ site.  This preference of the octahedral site, which has also been reported by several other groups \cite{GAO2020355,wei2010site,CONNETABLE201932307}, can be attributed to the larger volume of the octahedral site, which minimizes steric repulsion and lattice strain. Moreover, the higher coordination number of the octahedral site offers a more stable and energetically favorable environment compared to tetrahedral or lower co-ordination sites.

The presence of an interstitial dopant can significantly impact the formation energy of an alloy by altering its lattice structure and electronic properties, and to assess its effect on the overall stability of the material, the formation energies of the doped structures were calculated at different doping concentrations using Eq. \ref{eqn:E1}, as presented in Fig. \ref{figure:2}. Interestingly, the introduction of interstitial atoms resulted in enhanced structural stability compared to the pristine compound, as indicated by the negative formation energy $E_\mathrm{f}$. This indicates the solubility of the interstitial atoms in Fe$_2$MnSn and implies that the  formation of a single-crystalline structure is favorable rather than undergoing phase separation.  At lower interstitial concentrations, the entropy of the doped compounds increases due to the introduced  disorder, while the electronic and magnetic interactions are modified by the presence of the dopant. This enhances the stability of the crystal structure, potentially resulting in a large negative formation energy, as seen in Fig. \ref{figure:2}.  However, at higher doping concentrations, competition between different lattice formations arises, disrupting atomic arrangements and weakening bonding strength. This destabilizes the system, leading to a less stable structure with the formation energy becoming less negative as compared to the relatively lower doping concentration, as shown in Fig. \ref{figure:2}.

\subsection{Switch in magneto-crystalline anisotropy }
\label{subsec2}

\begin{figure}[t]
\centering
\includegraphics[scale=0.50]{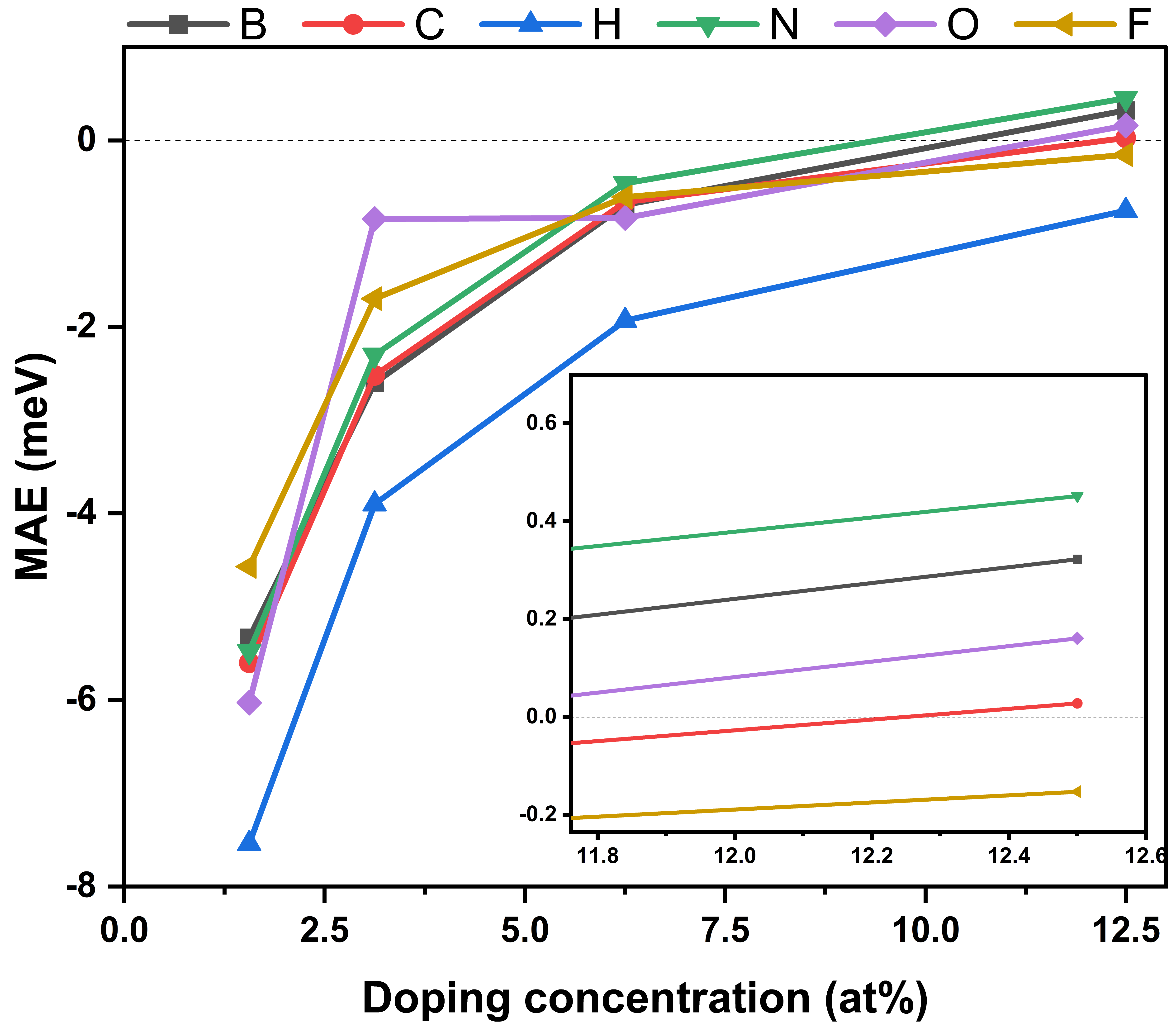}
\caption{Magneto-crystalline anisotropy energy (MAE) variation with increase in interstitial concentration of interstitial atoms. The inset highlights the magnetic anisotropy obtained for the 12.5 at\% interstitial doping concentration.}
\label{figure:3}
\end{figure}

A substantial magneto-crystalline anisotropy (MCA), marked by a positive magnetic anisotropy energy with an out-of-plane magnetization orientation is essential for PMs, as it governs the stability of the magnetization orientation and enhances the  magnetic coercivity, thereby providing greater resistance to demagnetization. MAE quantifies the energy required to reorient magnetization from in-plane to out-of-plane, as defined in Eq. \ref{eqn:E2}. As shown in Fig. \ref{figure:3}, the MAE of all the doped compounds remains predominantly in-plane at interstitial concentrations ranging from 1.56 at\% to 6.25 at\%, although it becomes progressively less negative with increased doping. Remarkably, at a concentration of 12.5 at\%, we observe a transition from in-plane to out-of-plane MAE in all interstitially modified compounds except for H-FMS and F-FMS. The shift in magnetic anisotropy can be attributed to doping-induced lattice distortions absent in H-FMS, which can alter the crystal-field, and modify the energy difference between different magnetic orientations \cite{PhysRevB.14.2287}. As also summarized in Table \ref{table:1}, N-FMS exhibits the highest MAE ($K$) with a value of 0.45 meV (0.61 MJ/m$^3$), followed by B-FMS (0.32 meV, 0.44 MJ/m$^3$). While C-FMS and O-FMS also display an out-of-plane anisotropy, their MAE values are significantly lower at 0.02 meV and 0.16 meV, respectively. Since achieving uniaxial magnetic anisotropy is the key objective of this study, noting that the undoped FMS exhibited an in-plane anisotropy with a negative MAE ($K$ = -1.24 MJ/m$^3$) \cite{PhysRevB.108.054431}, we decided to focus further investigations solely at the 12.5 at\% interstitial concentration for which an out-of-plane anisotropy was obtained for most of the dopants.

In order to explore the origin of induced MAE by interstitial atoms, we have employed perturbation theory, since the origin of MAE is primarily derived from SOC, and is relatively weak in ferromagnetic transition metals. We used Bruno's model \cite{PhysRevB.39.865} derived from second-order perturbation theory, which states that the MAE is proportional to the difference in orbital moments between the easy and hard axes, with the easy axis corresponding to the direction that maximizes the orbital moment. While originally developed for layered systems which have quite large surface MAE, we evaluated its validity for our bulk systems using Eq. \ref{eqn:E5}  by calculating the difference in orbital moments $\Delta \mu_{l} = \mu_{l} [001] - \mu_{l} [100]$ such that in case of uniaxial MAE, $\Delta \mu_{l}$ is positive.
\begin{equation}
	\label{eqn:E5}
	 \mathrm{MAE}=\frac{\xi}{4\mu_\mathrm{B}}\Delta {\mu_l}
\end{equation}
 where, $\xi$ is the SOC constant and $\mu_\mathrm{B}$ is Bohr's magneton.
 
\begin{figure}[t]
\centering
\includegraphics[scale=0.50]{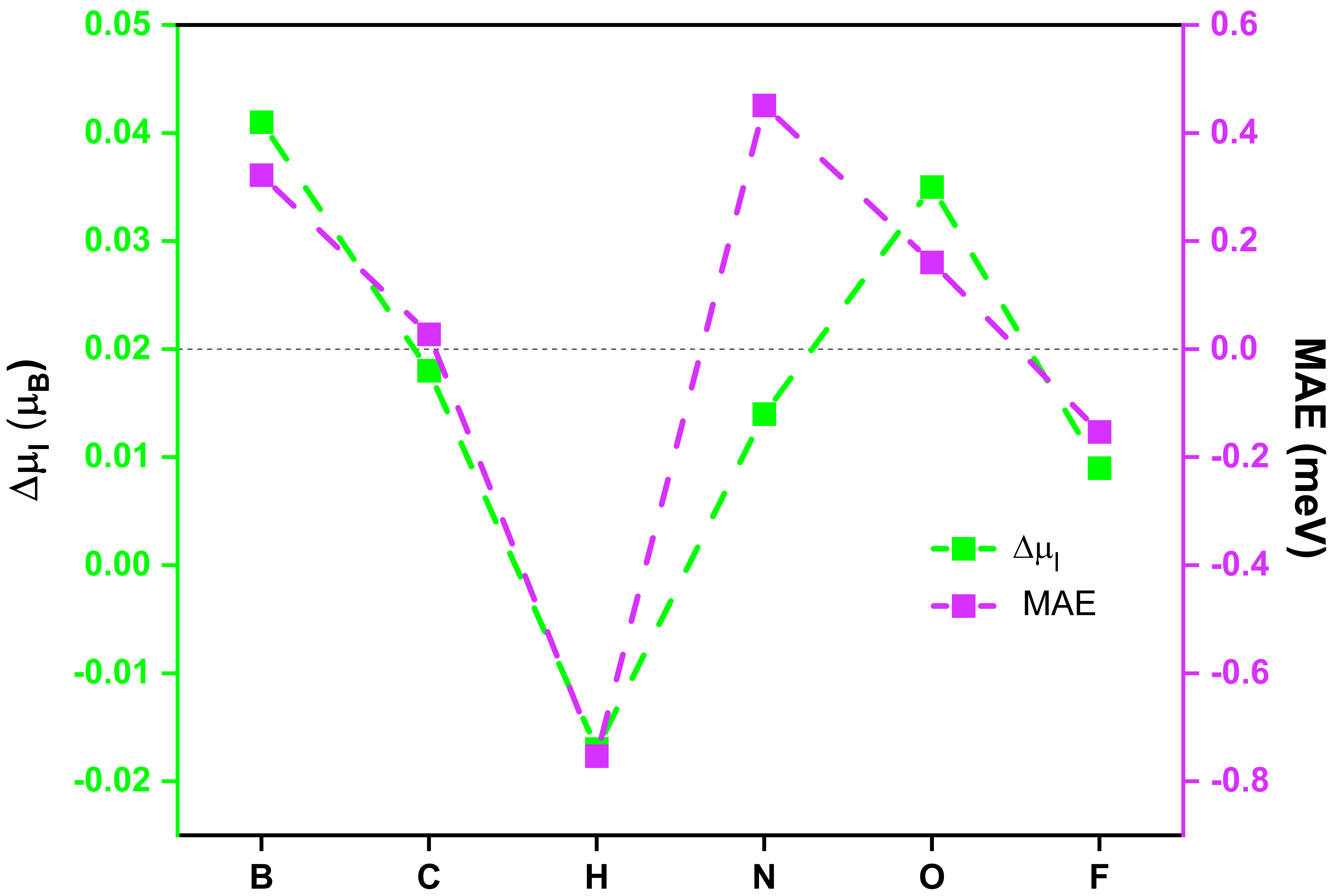}
\caption{The difference in orbital moments between the easy and hard axis ($\Delta \mu_{l}$ in green) and the corresponding magneto-crystalline anisotropy (MAE in magenta) for the 12.5 at\% interstitially modified Fe$_2$MnSn plotted for different interstitial dopants.}
\label{figure:4}
\end{figure}

 As illustrated in Fig. \ref{figure:4}, B-FMS shows the largest difference in orbital moments, suggesting it should have the highest MAE based on this model. However, our results reveal that the N-FMS actually demonstrates the largest MAE, indicating that additional factors must be taken into account,  highlighting the limitations of Bruno's model in this scenario. This discrepancy may arise from strong hybridization between the electronic states of different atomic species in a multi-component system. 
 Here, the simplified relation proposed by Bruno, considering only the spin-conserved terms, becomes more complex and cannot be directly applied, as the off-site spin-flip interactions, typically ignored in Bruno's relation, play a critical role in determining the overall MAE in such systems.
 
\begin{figure*}
\centering
\includegraphics[scale=0.30]{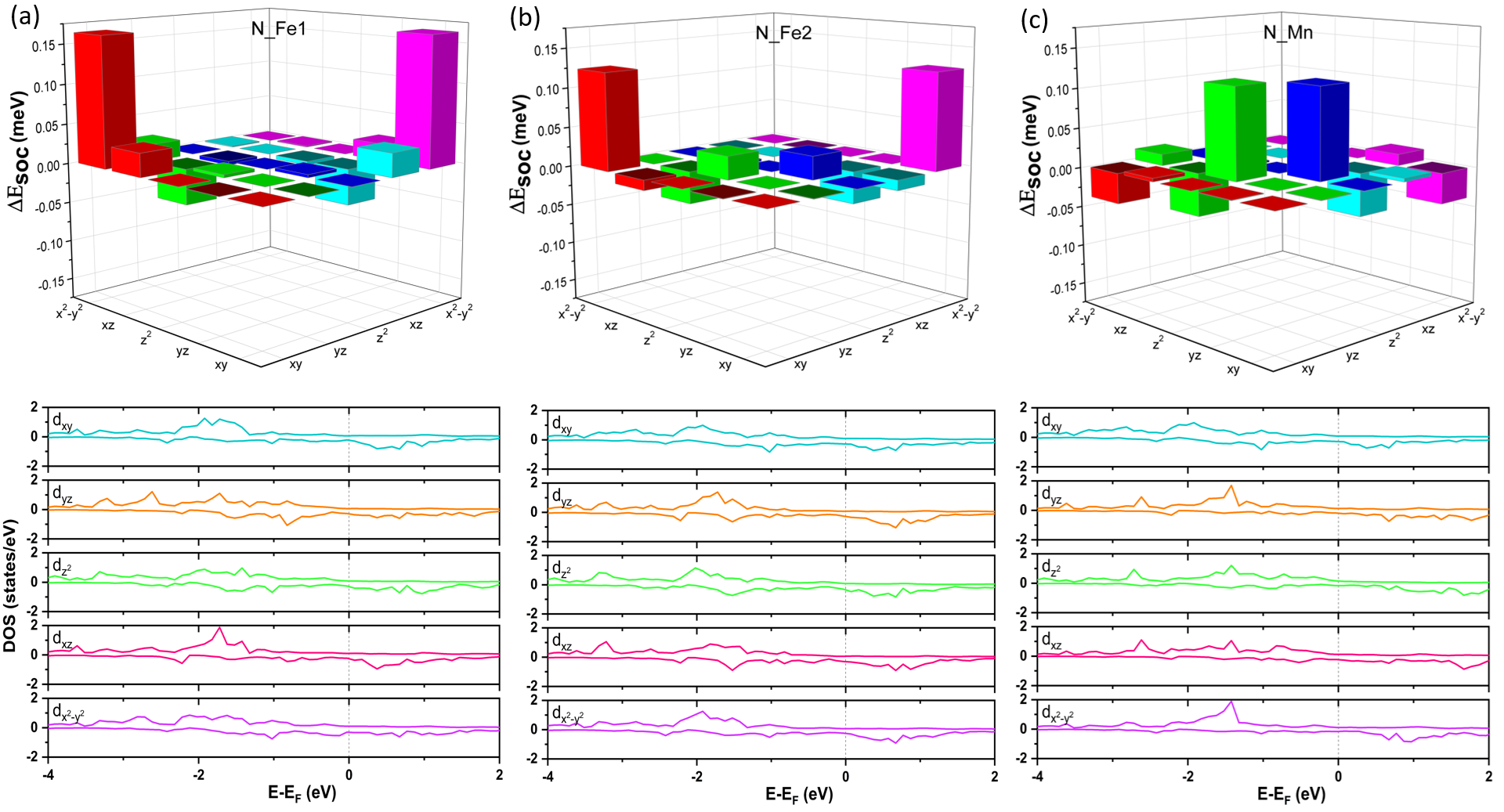}
\includegraphics[scale=0.30]{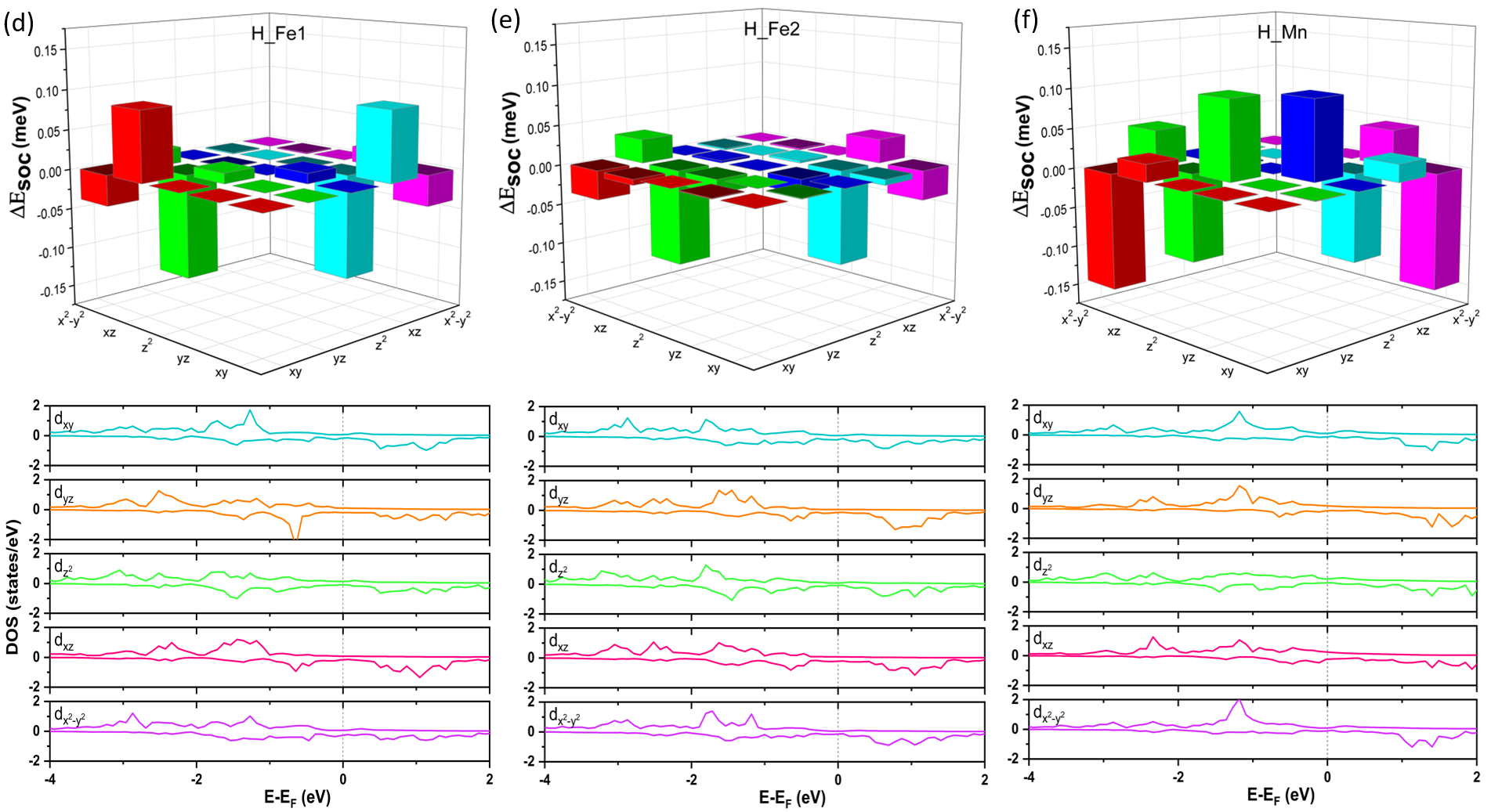}
\caption{The atom (Fe1, Fe2 and Mn) resolved
 contributions to MAE as calculated from their SOC interaction energy ($\Delta E \mathrm{_{SOC}}$) along with the projected $d$ orbital resolved DOS for two exemplary cases viz. the N-FMS [(a), (b), and (c)] and the H-FMS [(d), (e), and (f)].}
\label{figure:5}
\end{figure*}
	
In the framework of second-order perturbation theory, the energy correction due to SOC, which provides a quantitative value for the MAE, can be obtained using Eq. \ref {eqn:E6}, by accounting for both spin-conserved and spin-flip terms.  This approach allows for a more detailed understanding of how different atomic sites contribute to the overall anisotropy by examining the $d$-orbital projected contributions to the MAE for the symmetrically inequivalent Fe (i.e., Fe1 and Fe2) and Mn atoms. Half of the SOC energy as given by, $\hat{H}_\mathrm{SOC} = \xi \hat{L} \cdot \hat{S}$, contributes to the MAE i.e., $\mathrm{MAE} \approx \frac{1}{2}\Delta E_\mathrm{SOC}$ \cite{antropov2014constituents,PhysRevB.96.014435}, with the other half leading to an increase in crystal-field energy and formation of unquenched orbital moments \cite{skomski2011magnetic}.  
\begin{equation}
	\label{eqn:E6}
	\mathrm{MAE} = \xi^{2}\sum_{o\alpha,u\beta}^{ } (2\delta _{\alpha \beta }-1) \bigg[ \frac {\left|\bra {o\beta} L_{z}\ket{u\alpha}\right|^{2} - \left|\bra{o\beta}L_{x}\ket{u\alpha}\right|^2    }{E_{u\alpha} - E_{o\beta}  }\bigg] 
\end{equation}

where $\xi$ is the spin-orbit coupling strength, $\delta$ is the Kronecker function, $o\beta$ $(u\alpha)$ and $E_{o\beta}$ $(E_{u\alpha})$ stand for the eigen state and eigen value of occupied (unoccupied) orbital state $d_{xy}$, $d_{yz}$, $d_{z^2}$, $d_{xz}$, or $d_{x^2-y^2}$ with the up spin $\alpha$ and down spin $\beta$, respectively. Also, $L_z$ ($L_x$) denotes the $z$($x$) component of the orbital angular momentum operator.  The relative contributions of the non-zero SOC matrix elements of the $d$-orbital angular momentum operators are $\bra{d_{xy}}L_x\ket{d_{xz}} = 1$, $\bra{d_{x^2-y^2}}L_x\ket{d_{yz}} = 1$, $\bra{d_{z^2}}L_x\ket{d_{yz}} = \sqrt{3}$, $\bra{d_{xz}}L_z\ket{d_{yz}} = 1$, and $\bra{d_{x^2-y^2}}L_z\ket{d_{xy}} = 2$. The MAE (as expressed in Eq.~\ref{eqn:E6}) can be decomposed into two components based on spin orientations: parallel spin orientation MAE$^{\downarrow \downarrow}$ ($\alpha \beta = \uparrow \uparrow$ or $\downarrow\downarrow$) and  antiparallel spin orientation MAE$^{\uparrow \downarrow}$ (($\alpha \beta = \uparrow \downarrow$ or $\downarrow \uparrow$ ). For parallel spins, positive (negative) contributions to the MAE are provided by the  $L_z$ ($L_x$) matrix elements, while for anti-parallel spins ($\alpha \beta = \uparrow \downarrow$ or $\downarrow \uparrow$ ), the contributions are reversed. The states near the Fermi energy $E_\mathrm{F}$ contribute the most to the MAE, with the energy difference in the denominator of Eq. \ref{eqn:E6} affecting the magnitude of energy change, while the matrix elements in the numerator dictate the orientation of the MAE \cite{PhysRevB.98.014411}.

To further investigate the underlying factors behind the difference in magnetic anisotropies, we selected N-FMS and H-FMS for detailed analysis, as the former exhibits the highest uniaxial MAE, while the latter retains in-plane MAE even at 12.5 at\% interstitial concentration.  For both compounds as seen from Fig. \ref{figure:5}, the $d$-orbital-projected DOS of Fe and Mn atom reveals that near the Fermi level, the majority spin states are nearly fully occupied, whereas the minority spin states are partially filled. This suggests that the MAE contributions arise from occupied spin-up and spin-down $d$ states and unoccupied spin-down $d$ states. The $d$-orbital resolved contributions to the MAE for the other alloys namely B-FMS, C-FMS, O-FMS and F-FMS can be found in the supplementary information. 

For the N-doped compound, the matrix element $\bra{x^2-y^2\uparrow}L_z\ket{xy\uparrow}$ dominates the uniaxial MAE in both Fe atoms, indicating that ${d_{x^2-y^2}\uparrow}$ orbital couples through the same spin channel with ${d_{xy}\uparrow}$ orbital.  Hence, the occupation of $x^2-y^2$ state and the lack of occupation of $xy$ state will strongly favor out-of-plane anisotropy. Moreover, the $d$-orbital-projected DOS analysis reveals a notable distinction between the Fe1 and Fe2 atoms near the Fermi level. The energy difference between the spin-conserved occupied $d_{x^2-y^2}$ and unoccupied $d_{xy}$ states of Fe2 is slightly larger than that of Fe1 resulting in a higher magneto-crystalline anisotropy energy for Fe1, as the smaller energy gap in Fe1 leads to a reduced denominator term in Eq. \ref {eqn:E6} , thereby contributing to the larger MAE. Minor contributions arise from $\bra{xy\uparrow}L_x\ket{xz\downarrow}$ and  $\bra{z^2\uparrow}L_x\ket{yz\downarrow}$ for Fe1 and Fe2 respectively, resulting in total MAE contributions of 0.179 meV/atom and 0.106 meV/atom from the corresponding atoms. In the Mn atoms, the primary contribution to MAE originates from the SOC matrix element $\bra{z^2\uparrow}L_x\ket{yz\downarrow}$, where ${d_{z^2}\uparrow}$ couples via spin-flip channel with ${d_{yz}\downarrow}$. Additionally, small negative contributions arise from $\bra{xz\uparrow}L_z\ket{yz\downarrow}$, resulting in a modest total MAE contribution of 0.056 meV per Mn atom. In contrast, H-FMS compound exhibits an in-plane MAE with largest contribution coming from Fe2 atom (-0.137 meV), followed by Mn (-0.076 meV) and Fe1 (-0.033 meV), indicating that negative contribution to magnetic anisotropy is observed for all of the magnetic elements, forcing the compound to exhibit planar anisotropy. In both Fe1 and Fe2, the matrix element $\bra{xz\uparrow}L_z\ket{yz\downarrow}$ is significant, indicating that the anti-parallel coupling of the occupied $xz \uparrow$ and unoccupied $yz \downarrow$ orbitals drive planar anisotropy. Fe1 also receives a small positive contribution from $\bra{xy\uparrow}L_x\ket{xz\downarrow}$, resulting in a relatively weak net contribution. In Mn atom, $\bra{x^2-y^2\uparrow}L_z\ket{xy\downarrow}$ dominates the in-plane MAE, followed by $\bra{xz\uparrow}L_z\ket{yz\downarrow}$. However, there is also a positive contribution stemming from the anti-parallel coupling between the ${d_{z^2}\uparrow}$  and ${d_{yz}\downarrow}$ orbitals.

\begin{table}[!t]
	
	   \caption{Optimized lattice parameters $a$ and $c$ (\AA), formation energy $E_\mathrm{f}$ (eV/f.u.), magneto-crystalline anisotropy energy MAE (meV), anisotropy constant K (MJ/m$^3$), spin polarization SP, net spin magnetic moment M$_\mathrm{net}$ ($\mu_\mathrm{B}$/f.u.) stemming from individual uncompensated magnetic moments ($\mu_{\mathrm{Fe1}}$, $\mu_{\mathrm{Fe1}}$, $\mu_{\mathrm{Fe2}}$, $\mu_{\mathrm{Mn}}$, $\mu_{\mathrm{Sn}}$, and $\mu_{\mathrm{I}}$), volume expansion term $V_{\mathrm{exp}}$, saturation magnetization $\mu_0 M_s$ (T), maximum energy product BH$_{\mathrm{max}}$ (MJ/m$^3$), hardness parameters $\kappa$, and Curie temperature T$_{\mathrm{C}}$ (K) calculated for interstitially modified Fe$_2$MnSn Heusler alloys. For all the doped compounds, the lattice parameters $b$ and $c$ are observed to be equal, maintaining the inherent symmetry of the structure. However, in the case of F-FMS, this symmetry is disrupted, with $b$ and $c$ becoming unequal. Notably, $b$ is significantly reduced to 5.32 \AA, indicating a pronounced structural distortion.}
    \label{table:1} 
	\begin{tabular}{p{2.3cm}p{2cm}p{2cm}p{2cm}p{2cm}p{2cm}p{1.5cm}} 
		\hline
	
		Parameters& B & C & H & N & O & F \\ 
		\hline
		{$a$}& 5.48 & 5.52 & 5.45 & 5.56 & 5.59 & 6.33 \\
		{$c$}& 4.53 & 4.44 & 4.40 & 4.41 & 4.52 & 4.92 \\
		{$E_\mathrm{f}$} & -0.183 & -0.017 & -0.089 & -0.268 & -0.971 & -0.626 \\
		{MAE} & 0.32 & 0.02 & -0.75 & 0.45 & 0.16 & -0.15\\ 
		{K} & 0.437 & 0.037 & -1.063 & 0.611 & 0.210 & -0.185 \\
		{SP} & 22 & 25 & 14 & 36 & 18 & 30 \\
		{M$_{\mathrm{net}}$} & 6.41 & 6.29 & 6.66 & 6.78 & 7.85 & 7.93 \\
		{$\mu_{\mathrm{Fe1}}$}& 1.986 & 1.978 & 2.061 & 2.067 & 2.430 & 2.484 \\
		{$\mu_{\mathrm{Fe2}}$}& 2.085 & 2.037 & 2.215 & 2.195 & 2.490 & 2.522 \\
		{$\mu_{\mathrm{Mn}}$}& 2.432 & 2.373 & 2.366 & 2.544 & 2.848 & 3.074 \\
		{$\mu_{\mathrm{Sn}}$}& -0.141 & -0.145 & -0.154 & -0.144 & -0.150 & -0.186 \\
		{$\mu_{\mathrm{I}}$}& -0.197 & -0.173 & -0.024 & -0.104 & 0.127 & 0.068 \\
		{$V_{\mathrm{exp}}$} & 5.85 & 4.75 & 1.70 & 5.93 & 10.10 & 18.40 \\
		{$\mu_0M_s$} & 1.27 & 1.26 & 1.37 & 1.34 & 1.49 & 1.40 \\
		{BH$_{\mathrm{max}}$} & 0.32 & 0.31 & -- & 0.36 & 0.44 & -- \\
		{$\kappa$} & 0.59 & 0.17 & -- & 0.65 & 0.34 & -- \\
		{T$_{\mathrm{C}}$} & 986 & 1000 & 905 & 744 & 1058 & 703 \\
		
		\hline
		
	\end{tabular}
 
\end{table}

\subsection{Magnetization and Hardness Parameter}
\label{subsec3}
\begin{figure}[t]
\centering
\includegraphics[scale=0.50]{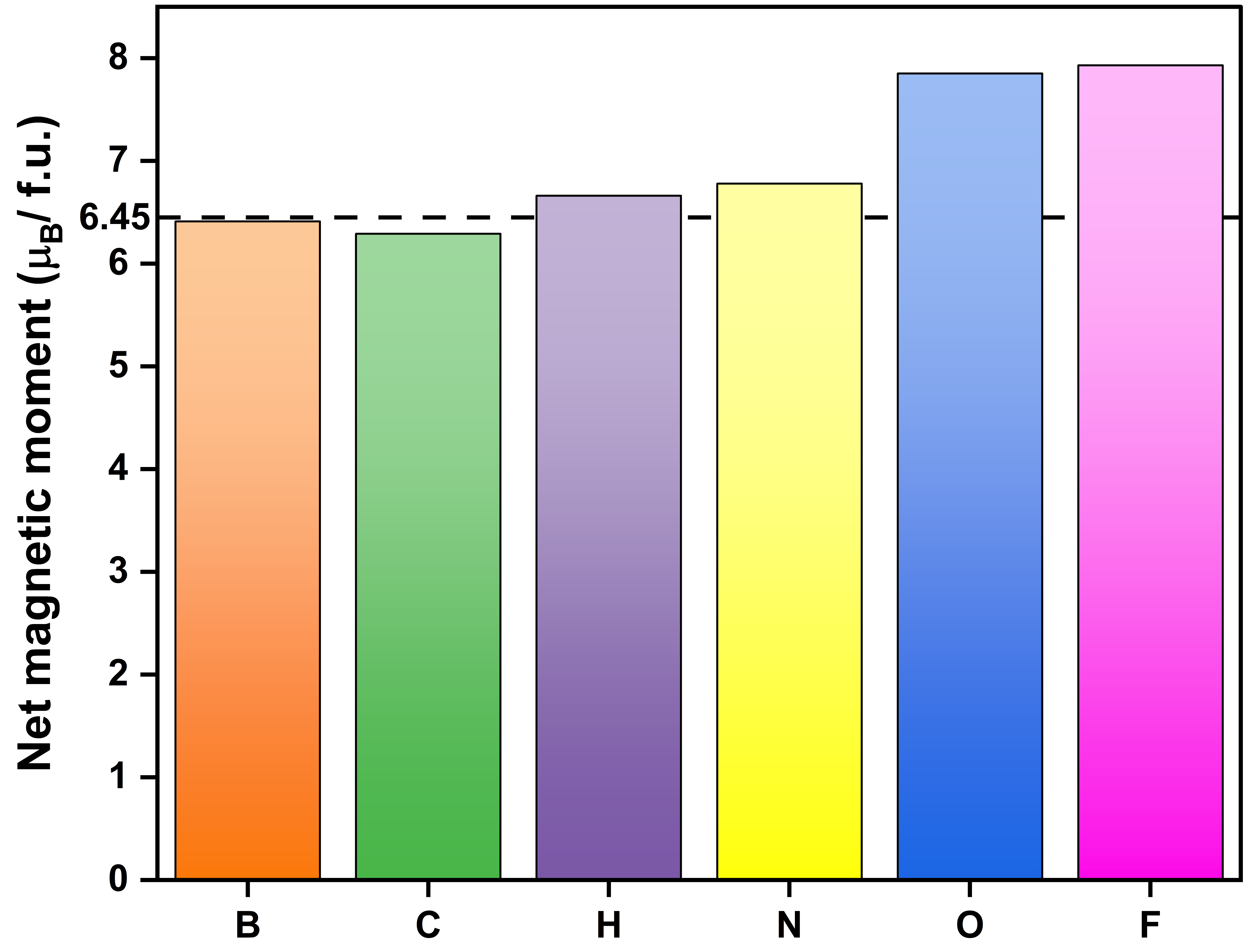}
\caption{Net spin magnetic moment for 12.5 at\% interstitially modified Fe$_2$MnSn. The horizontal dashed line at 6.45 $\mu_\mathrm{B}$/f.u. marks the magnetic moment of pristine Fe$_2$MnSn, thereby showing that interstitial doping lead to an increase in magnetization in most cases.}
\label{figure:6}
\end{figure}

Our findings demonstrate that the incorporation of interstitial atoms  significantly alters the magnetic properties of Fe$_2$MnSn Heusler alloy. In this $d$-orbital dominated alloys, the metallic bonding undergoes notable changes due to orbital mixing between the $s–p$ orbitals of the dopant and the $s–p–d$ orbitals of the host metals$-$an interaction absent in the pristine metal$-$resulting in variations in electronic and magnetic properties.  Increased interstitial doping (12.5 at\%) enhances $p-d$ orbital hybridization between magnetic and interstitial atoms, altering the electronic density of states (DOS) near the Fermi level as seen from Fig. S3 in supplementary information,  and directly influences the total magnetic moment of the interstitially modified compounds, as depicted in Fig. \ref{figure:6}. The asymmetry in the distribution of spin-up and spin-down states at the Fermi level (E$_\mathrm{F}$) due to the  exchange splitting, is consistent with the ferromagnetic behavior of these alloys.

When examining the total magnetic moment, the doped samples show a substantial increase in the net spin magnetic moment for H, N, O and F-FMS as detailed in Table \ref{table:1}, compared to the undoped compound (6.45 $\mu_\mathrm{B}$/f.u.) \cite{PhysRevB.108.054431}. Specifically, the introduction of oxygen and fluorine leads to more pronounced differences between the spin-up and spin-down states than other dopants, which directly contributes to the larger total moments in these compounds. This increase is further explained by the substantial fractional change in volume, calculated as ${V_{\mathrm{exp}}=\Delta V/{V_{\mathrm{undoped}}}} \times 100$, where $\Delta V = V_{\mathrm{doped}}-V_{\mathrm{undoped}}$. As the lattice expands, the orbital overlap decreases, giving rise to weaker hybridization and reduction of itinerancy, which allows a more atom-like (localized) behavior of the moments, hence making it larger. This effect aligns with the Bethe-Slater curve, where larger distances favor ferromagnetic coupling. A nearly monotonic increase in the magnetic moment is observed as $V_{\mathrm{exp}}$ increases, with the exception of H-FMS as can be seen in section S3 of supplementary information.

For the boron and carbon-doped compounds, the magnetic moment slightly decreases compared to the pristine compound (6.45 $\mu_\mathrm{B}$/f.u.), despite the accompanying volume expansion. This suggests that there has to be another competing phenomenon working in tandem that affects the net magnetization of the doped compounds. First, volume expansion generally enhances the magnetic moment by increasing the interatomic distances, which reduces orbital overlap and localizes the magnetic moments on individual atoms. Conversely, strong hybridization or bonding between the interstitial atom and nearby magnetic atoms can counteract this by delocalizing the magnetic moments, reducing their individual contributions and weakening the overall magnetization. The interplay between these opposing effects ultimately dictates the net change in magnetic moment in interstitially modified systems. To further investigate the strength of the chemical bonding between magnetic atoms (Fe and Mn) nearest to the interstitial atom, we calculated the spin-polarized integrated crystal orbital Hamilton population (ICOHP) values. The interactions between Fe-B and Fe-C show strong bonding with ICOHP values of -2.70 eV and -2.81 eV respectively. In contrast, the interactions involving hydrogen (Fe-H), nitrogen (Fe-N), oxygen (Fe-O), and fluorine (Fe-F) exhibit much lower ICOHP values of -0.77 eV, -2.33 eV, -1.34 eV, and -0.006 eV, respectively, which signify weaker bonding in these cases. Moreover, the ICOHP values for manganese (Mn) with boron and carbon are also noteworthy; Mn-B has an ICOHP of -27 meV, while Mn-C has an ICOHP of -26 meV. These values are nearly double those of Mn-O, which has an ICOHP of -14 meV, while Mn-H exhibits an ICOHP of -16 meV. This observation further underscores the significantly stronger bonding characteristics present in the systems doped with boron and carbon compared to those with oxygen and hydrogen. The strong hybridization in the B and C-doped systems appears to dominate over the volume expansion effect, ultimately leading to a reduction in the magnetic moment as reflected in the data presented in the Table \ref{table:1}. An increase in the net magnetic moment of H-FMS is also observed, despite the small volume expansion which can be attributed to hydrogen’s lower electronegativity and the fact that hydrogen is lighter, compared to boron and carbon. As a lighter element, hydrogen induces minimal perturbations in the spin density of neighboring atoms, which allows for an enhanced magnetic moment contribution from the metal atoms. In contrast, boron’s and carbon's stronger hybridization  with the metal atoms evident from the ICOHP values, likely reduces the localized magnetic moments on these atoms, even though it results in an increased volume. Overall, the computed total magnetization for doped FMS (B = 1.27 T, C = 1.26 T, H = 1.37 T, N = 1.34 T, O = 1.49 T and F = 1.40 T) surpass those of conventional hard ferrites ($\sim$ 0.40 T) \cite{MOHAPATRA20181, SKOMSKI20163}. These values also exceed the magnetization of widely studied gap magnets such as MnAl (0.82 T) \cite{pareti1986magnetic,wei2014tau} and MnBi (0.8$-$0.93 T) \cite{PhysRevB.83.024415,YANG2021157312,guo1992magnetic}, and are comparable to that of FeNi alloys (1.5$-$1.72 T) \cite{tuvshin2021first,jacob2021enhanced,liu2019magnetic,TUVSHIN2021116807}.

An estimation of the potential of these materials for permanent magnets can be made by evaluating key intrinsic hard magnetic properties. One such property is the maximum energy product, BH$_{\mathrm{max}}$ = $\mu_0 M_s{^2}/4$, which represents the theoretical upper limit assuming an ideal rectangular $M-H$ loop. Here, $\mu_0$ is the vacuum permeability, $M_s$ is the saturation magnetization of the material. A significant energy product is observed for all doped compounds, with values ranging from 0.31$-$0.44 MJ/m$^3$, closely comparable to the well known rare earth neo-magnets (0.36$-$0.4 MJ/m$^3$) \cite{honshima1994high,scott1996microstructural}. Notably, O-FMS shows the highest energy product ($\sim$ 0.44 MJ/m$^3$), as presented in Table \ref{table:1}. Another important property is the magnetic hardness parameter, $\kappa$   =$\sqrt{\frac{K}{\mu_0 M_s^{2}}}$ \cite{1130282271085861760,6008648}, a dimensionless quantity  used to classify ferromagnetic materials  as hard, semi-hard, or soft. A value of $\kappa\geq1$ represents hard magnets, and $\kappa \geq 0.1$ indicate semi-hard magnets. Permanent magnets fall within the hard or semi-hard range with $\kappa > 0.1$. For reference the well known permanent magnets viz. Nd$_2$Fe$_{14}$B and SmCo$_5$ exhibit $\kappa$ values of 1.54 and 4.4, respectively, while AlNiCo and FeNi alloys have  $\kappa$ values around 0.4$-$0.5  \cite{OCHIRKHUYAG2024119755,SKOMSKI20163}. In our study, N-FMS displays the highest $\kappa$ (0.65), followed by B-FMS (0.59), indicating their potential as gap magnets.

 \begin{figure}[t]
\centering
\includegraphics[scale=0.50]{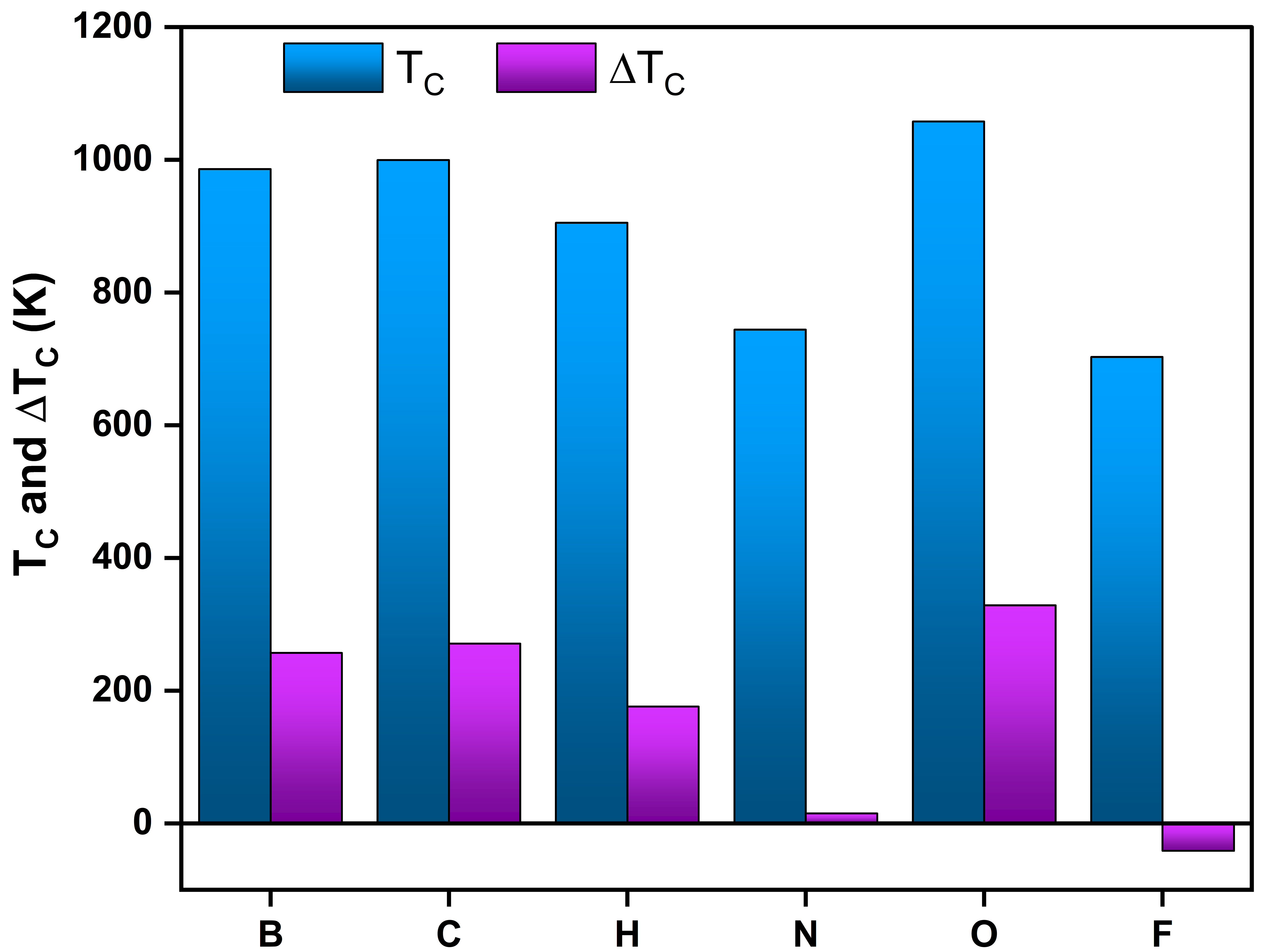}
\caption{Curie temperature T$_{\mathrm{C}}$ and the change incurred in comparison to the T$_{\mathrm{C}}$ of the pristine alloy $\Delta$T$_{\mathrm{C}}$ for the 12.5 at\% interstitially modified Fe$_2$MnSn. For most of the interstitial dopants, the T$_{\mathrm{C}}$ was found to either increase or comparable to the pristine alloy.}
\label{figure:7}
\end{figure}

\subsection{Curie Temperature and Exchange Interactions}
\label{subsec4}
For a magnet to be practical, it must maintain its magnetic properties under operational conditions, when exposed to elevated temperatures. The Curie temperature (T$_{\mathrm{C}}$) marks the point where there is no spontaneous bulk magnetization and the long-range magnetic ordering vanishes, due to a magnetic order-disorder transition. In applications such as magnetic sensors in aeroengines or permanent magnets in electric motors, T$_{\mathrm{C}}$ must significantly exceed room temperature \cite{brannvall2024predicting}.   In all interstitially modified Fe$_2$MnSn Heusler alloys considered in our study, an enhancement in T$_{\mathrm{C}}$ is observed relative to the pristine alloy (729 K)\cite{PhysRevB.108.054431}, with the exception of F-FMS (see Fig. \ref{figure:7}). This indicates the positive effect of interstitial atoms on magnetic stability. Among the doped compounds, O-FMS achieves the highest T$_{\mathrm{C}}$ of 1058 K (which is 45\% higher than the pristine alloy), closely followed by C-FMS at 1000 K. Despite showing the smallest increase of only 2\% (i.e., 15 K), N-FMS still achieves a relatively high Curie temperature of 744 K. In contrast, F-FMS shows a slight decline in T$_{\mathrm{C}}$, reaching 703 K, representing a reduction of 26 K, as illustrated in Fig. \ref{figure:7} and highlights the dependence of T$_{\mathrm{C}}$ on the dopant involved.   Theoretically, mean-field approximation has been found to generally overestimate T$_{\mathrm{C}}$ \cite{PhysRevB.64.174402}. Nevertheless, our computed T$_{\mathrm{C}}$ meets the essential condition of high-performance PMs having T$_{\mathrm{C}}$ greater than 550 K, as proposed by Coey \cite{COEY2012524}. 

 \begin{figure}[t]
\centering
\includegraphics[scale=0.30]{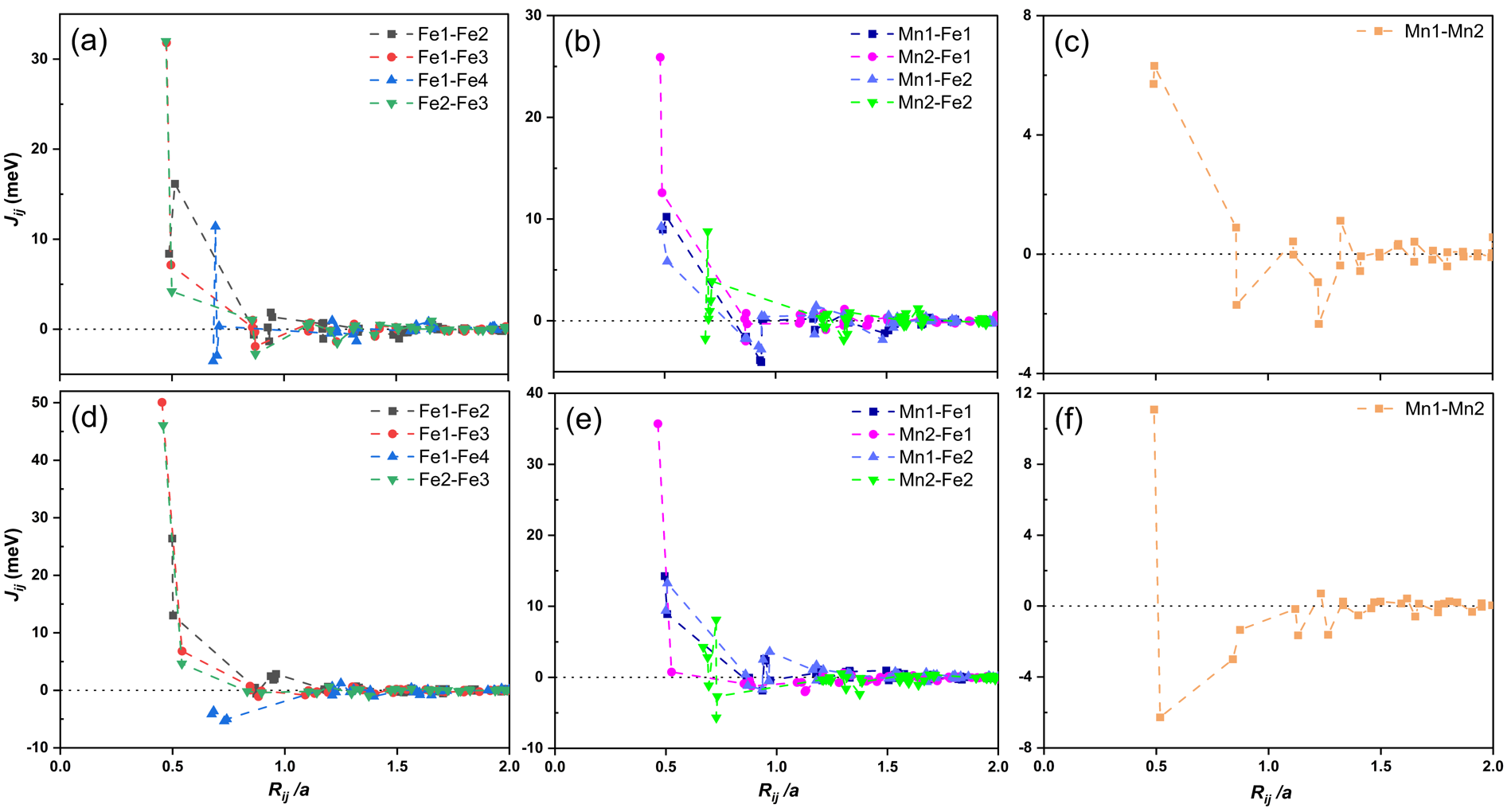}
\caption{Heisenberg exchange coupling parameters $J_{ij}$ plotted as a function of the interatomic distance scaled by the lattice parameter $a$ ($R_{ij}/a$) for N-FMS (a), (b), (c) and O-FMS (d), (e), (f).}
\label{figure:8}
\end{figure}

 The observed enhancement in T$_{\mathrm{C}}$ can be linked to the strengthening of the ferromagnetic coupling between magnetic atoms (Fe and Mn) following interstitial modification and can be better understood with the help of Bethe-Slater curve  which helps explain the relationship between interatomic distances and magnetic interactions \cite{cardias2017bethe}.  On the Bethe-Slater curve, Fe lies in the FM region, near the FM-AFM boundary, hence shorter Fe–Fe distances promote an AFM state, while increased Fe–Fe distances reinforce FM coupling and thereby raising T$_{\mathrm{C}}$ \cite{kitagawa2020interstitial}. Similarly, Mn, which typically exhibits an AFM ground state on the Bethe-Slater curve, can transition toward FM behavior as Mn–Mn distances increase. We see for all doped compounds, unit cell expansion is observed (Table \ref{table:1}), leading to an increase in the Fe-Fe and Mn-Mn interatomic distances and hence strengthening of FM coupling, explaining the rise in T$_{\mathrm{C}}$ across the modified compounds. However, the Bethe-Slater model falls short in explaining certain anomalies observed in our results. For instance, despite having a lattice expansion comparable to C-FMS and O-FMS, N-FMS exhibits a lower T$_{\mathrm{C}}$, which contradicts the expected trend based on the curve. Likewise, F-FMS, which undergoes the most significant lattice expansion, paradoxically shows a reduction in T$_{\mathrm{C}}$ instead of the anticipated enhancement in magnetic interactions. It should be noted that T$_{\mathrm{C}}$ is directly dependent on the exchange coupling parameters $J_{ij}$, which is oversimplified  when considering just the distance between magnetic atoms. The curve is based on nearest-neighbor interactions (direct interactions) and does not consider long-range magnetic interactions and primarily applies to pure transition metals and struggles to predict the magnetic behavior in alloys or compounds with multiple elements.  
 
To gain a deeper understanding of the variations in T$_{\mathrm{C}}$ with different interstitial atoms, we analyze the exchange coupling constants $J_{ij}$ for N-FMS and O-FMS, which exhibit the smallest and largest increase in T$_{\mathrm{C}}$, respectively. The exchange parameters for the other compounds (B-FMS, C-FMS, H-FMS, and F-FMS) are provided in section S4 of the supplementary information.  Assuming Heisenberg-type interaction, the exchange coupling parameters between the magnetic atoms are calculated using the Green's function method based on multiple scattering formalism \cite{kordt2012single}.  Positive exchange parameters indicate a preference for ferromagnetic coupling, where the spins of two magnetic moments align parallel to each other. Conversely, negative exchange parameters reflect a tendency for antiferromagnetic coupling, where the spins align antiparallel.  It is important to note that the system studied in this work contains atoms with partially filled $d$-orbitals, leading to long-range exchange interactions that extend over several lattice constants.  Fig. \ref{figure:8} presents the Heisenberg exchange coupling parameters obtained from our calculations performed for a cluster with a radius of 7$a$ (where $a$ is the optimized lattice parameter). Minimal changes were observed as the cluster radius increased beyond 2$a$, indicating that the calculations have converged sufficiently.

As expected, the nearest-neighbor interactions are the strongest in magnitude and positive, indicating FM coupling. The Fe-Fe interactions are particularly dominant considering the first neighbor, reaching maximum strength of 32 meV for N-FMS and approximately 50 meV for O-FMS, primarily driven by the overlapping 3$d$ orbitals as illustrated in Fig. \ref{figure:8} (a) and Fig. \ref{figure:8} (d). Beyond the second nearest neighbors, the interaction strength drops significantly. In the pristine FMS alloy, the Fe2-Fe3 interactions exhibit antiferromagnetic coupling \cite{PhysRevB.108.054431}, but upon interstitial modification these interactions transition to ferromagnetic due to local atomic distortions, contributing to increase in T$_{\mathrm{C}}$. Additionally, the Mn2-Fe1 interactions show a strong ferromagnetic coupling, particularly for O-FMS, where the strength reaches nearly 36 meV, while N-FMS shows a reduced strength by approximately 10 meV (cf. Fig. \ref{figure:8} (b) and (e)). For Mn-Mn interactions, which were strongly AFM in the pristine compound (-36 meV for first neighbors),  flips to FM upon doping evident from Fig. \ref{figure:8} (e) and (f), reaching 6 meV for N-FMS and 11 meV for O-FMS. However, the Mn-Mn interactions display alternating signs depending on the distance, suggesting a competition between FM and AFM coupling, which could influence the temperature dependence of the magnetic ordering. The overall net FM ordering in these compounds is primarily driven by the strong Fe-Fe interactions, with Fe-Mn interactions also contributing, albeit to a lesser extent. The interactions between magnetic and non magnetic atoms such as the Fe-Sn and Mn-Sn, are found to be very weak and nearly zero for all distances. The observed rise in T$_{\mathrm{C}}$ can likely be attributed to the enhanced ferromagnetic exchange coupling, modulated by the introduction of suitable interstitial atoms, thereby stabilizing the magnetic order.

\section{Conclusion}
\label{sec4}
The growing demand for novel permanent magnets, driven by renewable energy and zero-emission transport technologies, underscores the need for materials with enhanced magnetic properties. In this study, we have explored the potential of Fe$_2$MnSn Heusler alloy as a rare-earth free permanent magnet, using density functional theory calculations, through interstitial modification. The hexagonal ferromagnetic phase of Fe$_2$MnSn, identified as the ground state in our previous study, exhibits high saturation magnetization and large Curie temperature. However, its in-plane magnetic anisotropy limits its application, and a perpendicular (out-of-plane) anisotropy is preferred for permanent magnet use. With a focus to engineer an uniaxial magnetic anisotropy, we doped Fe$_2$MnSn with  six different interstitial atoms— B, C, H, N, O, and F—individually, at concentrations of 1.56, 3.125, 6.25, and 12.5 at\%. 
A detailed analysis of site preferences reveals that the octahedral $2a$ (0, 0, 0) site is favored for most dopants, except for fluorine, which preferentially occupies the octahedral $6g$ (0, 0.5, 0) site. At lower concentrations (1.56–6.25 at\%), the magneto-crystalline anisotropy energy remains in-plane. However, at 12.5 at\%, a transition from in-plane to out-of-plane MAE is observed in all doped compounds except H-FMS and F-FMS. This shift, driven by lattice deformation and the reorientation of the easy axis, was achieved without the need for 5$d$ or rare-earth elements, demonstrating potential of this alloy for rare-earth free permanent  magnet applications. Additionally, interstitial doping stabilizes the crystal structure and enhances both the magnetization and Curie temperature. The total magnetic moments of the doped compounds show a significant increase compared to the pristine alloy (6.45 $\mu_\mathrm{B}$/f.u.), with H, N, O, and F-doped systems showing notable improvements. The magnetization of the doped compounds (B = 1.27 T, C = 1.26 T, H = 1.37 T, N = 1.34 T, O = 1.49 T, and F = 1.40 T) exceeds those of conventional ferrites and gap magnets like MnAl and MnBi. All doped compounds exhibit significant energy products, ranging from 0.31–0.44 MJ/m$^3$, comparable to established rare-earth neo-magnets. N-FMS showed the highest magnetic hardness parameter (0.65), followed by B-FMS (0.59), confirming their potential as gap magnets. Among the dopants, O-FMS achieved the highest Curie temperature (1058 K), followed by C-FMS (1000 K). Despite a modest increase of only 15 K, N-FMS still reaches a respectable T$_{\mathrm{C}}$ of 744 K. The net ferromagnetic ordering in these compounds is primarily driven by strong Fe-Fe interactions, with a secondary contribution from Fe-Mn interactions. These findings highlight the potential of interstitial doping in Fe$_2$MnSn to develop gap magnets with improved magnetic properties, offering a promising pathway for permanent magnet applications.

\section{Declaration of Competing Interest}

The authors declare that they have no known competing financial interests or personal relationships that could have appeared to influence the work reported in this paper.

 \section{CRediT authorship contribution statement}
 \label{sec5}
\textbf{Junaid Jami:} Conceptualization, Investigation, Data curation, Methodology, Writing-Original Draft, Project administration. \textbf{Rohit Pathak:} Project administration, Visualization, Writing - Review \& Editing. \textbf{N. Venkataramani:} Writing - Review \& Editing, Supervision, Project administration. \textbf{K.G. Suresh:} Writing - Review \& Editing, Supervision, Project administration. \textbf{Amrita Bhattacharya:}   Supervision, Project administration, Writing - Original Draft, Writing - Review \& Editing, Funding acquisition.

\section{Acknowledgments}
\label{sec6}

 The high performance computational facilities viz. Aron (AbCMS lab, IITB), Dendrite (MEMS dept., IITB), Spacetime, IITB and CDAC Pune (Param Yuva-II) are acknowledged for providing the computational hours. JJ acknowledges funding received through PMRF grant (PMRF ID : 1300138). AB thanks funding received through BRNS regular grant (BRNS/37098) and SERB power grant (SPG/2021/003874). 

 \section{Declaration of Generative AI and AI-assisted technologies in the writing process}
 \label{sec7}
 During the preparation of this work the author(s) used ChatGPT for language refinement and improving manuscript readability. After using this tool/service, the author(s) reviewed and edited the content as needed and take(s) full responsibility for the content of the publication.








\end{document}